\documentclass[aps,pre,preprint]{revtex4-1} 
\pdfoutput=1
\usepackage{booktabs}
\usepackage{tabularx}
\usepackage{graphicx}
\usepackage{subcaption} 
\usepackage{dcolumn}
\usepackage{bm}
\usepackage{multirow} 
\usepackage{epstopdf} 
\usepackage{amsfonts}
\usepackage{amsmath}
\usepackage{amsthm}
\usepackage{setspace}
\usepackage[titletoc,toc]{appendix}
\usepackage{xcolor}
\usepackage{hyperref}
\usepackage{physics}	  
\hypersetup{colorlinks=true,linkcolor=blue,citecolor=blue,breaklinks={true}}
\usepackage[draft]{changes} 
\usepackage{enumerate}

\DeclareMathAlphabet\mathbfcal{OMS}{cmsy}{b}{n}

\graphicspath{{./Figures/}}
\DeclareGraphicsExtensions{.eps,.png,.jpg,.jpeg,.bmp,.gif,.pdf}

\theoremstyle{remark}
\newtheorem{remark}{Remark}   

\begin{document}

\title{Navigating differential structures in complex networks}

\author{Leonardo L. Portes}
\email{ll.portes@gmail.com}
\affiliation{Complex Systems Group, Department of Mathematics and Statistics, University of Western Australia, Nedlands, Perth, WA 6009, Australia}


\author{Michael Small}
\affiliation{Complex Systems Group, Department of Mathematics and Statistics, University of Western Australia, Nedlands, Perth, WA 6009, Australia}
\affiliation{Mineral Resources, CSIRO, Kensington, Perth, WA 6151, Australia}

\date{\today}

\begin{abstract}
Structural changes in a network representation of a system,  due to different experimental conditions or to its time evolution, can provide insight on its organization, function and on how it responds to external perturbations. 
The deeper understanding of how gene networks cope with diseases and treatments is maybe the most incisive demonstration of the gains obtained through this {\em differential} network analysis point-of-view, which lead to an explosion of new numeric techniques in the last decade.  
However, {\it where} to focus ones attention, or how to {\it navigate} through the differential structures in the context of large networks can be overwhelming even for few experimental conditions. 
In this paper, we propose a theory and a methodological implementation for the characterization of shared ``structural roles" of nodes {\em simultaneously} within and between networks, whose outcome is a highly {\em interpretable} map. 
The main features and accuracy are investigated with numerical benchmarks generated by a stochastic block model. 
Results show that it can provide nuanced and interpretable information in scenarios with very different (i) community sizes and (ii) total number of communities, and (iii) even for a large number of 100 networks been compared (e.g., for 100 different experimental conditions). 
Then, we show evidence that the strength of the method is its ``story-telling"-like characterization of the information encoded in a set of networks, which can be used to pinpoint unexpected differential structures, leading to further investigations and providing new insights. 
We provide an illustrative, exploratory analysis of four gene co-expression networks from two cell types $\times$ two treatments (interferon-$\beta$ stimulated or control). 
The method proposed here allowed us to elaborate a set of very specific hypotheses related to {\em unique} and {\em subtle} nuances of the structural differences between these networks --- which were then tested and confirmed in the original dataset. 
Finally, the method is flexible to address different research-field specific questions, by {\it not} restricting what scientific-meaningful characteristic (or relevant feature) of a node shall be used. 

\end{abstract}


\maketitle




\section{Introduction}
\label{sec.intro}
Modeling a complex system through a network representation, and examining the community structures that change (or on the other hand, remain coherent) along with different experimental conditions, can bring relevant information about the structure and function of the system's interacting parts. 
This {\it differential} network analysis is a very recent topic, which has been mainly developed by the bioinformatics community in the context of gene expression analysis (see \cite{Chowdhury2019} for a review). 
That approach has contributed to several discoveries related to how genes communicate to support the emergence of physiological responses of an organism in different disease states~\cite{Sonawane2019,VanDam2017,Anglani2014,Amar2013}.  
However, besides the existence of several numeric techniques to accomplish the task, {\it where} to focus ones attention on, or how to {\it navigate} through the differential structures in the context of large networks can be overwhelming even along with few experimental conditions.

In this paper we provide a method to unfold that information, allowing one to pinpoint {\it how} and {\it what} communities structures change (or remain the same) along with different conditions. 
Our approach is inspired by recent advances in the field of chaotic phase synchronization (PS) and through regarding its characterization and detection via multivariate-singular spectrum analysis (M-SSA).  
It has been shown that orthogonal-rotations of the eigenvectors, obtained from concatenated trajectory matrices, provide a clear and almost automatic identification of oscillatory modes that are being shared by the coupled chaotic oscillators~\cite{Portes2019,Portes2016a,Portes2016b,Groth2011}.  
Specifically, the final outcome in that version of M-SSA is a set of rotated eigenvectors that clearly encode the {\it shared oscillatory components} of those oscillators, which can be used for further detection of PS and characterization of phase synchronized clusters. 

Here, we make a parallel between shared oscillatory modes in PS analysis and the shared ``structural role" of nodes in differential network analysis. 
In short, our method takes advantage of (i) a varimax rotation to simplify the structure of the eigenvectors obtained from (ii) the eigendecomposition of (iii) a concatenated adjacency matrix that represents the network in different conditions. 
For the sake of discussion, we will call it the concatenate-decompose-rotate (CDR) approach. 
In the new context of network analysis, we show that the outcome is a highly {\it interpretable map} of the nuances of the network in those different conditions.

It is worth noting that some combination of those three steps (concatenate, decompose, rotate) has already been applied in other works and by different scientific communities~\cite{Xiao2014,Vejmelka2010}. 
However, both their main goals and outcomes are different from the CDR as articulated here.  
Indeed, we expect that the fundamental underlying ideas of the CDR could be applied over and above the other methods that have been developed within different scientific communities, or even be used as a bridge between them to nurture a deeper theoretical understanding --- 
for example, as the recently demonstrated equivalence of modularity maximization and the method of maximum likelihood~\cite{Newman2016}).

Nevertheless, there has been some criticism regarding the methodological aspects of community detection in network science~\cite{Fortunato2016}. 
In particular,  we share the pertinent view that a ``{\it method based on a mere hunch that something might work is inherently less trustworthy than one based on a provable result or fundamental mathematical insight.}"~\cite{Ball2011}. 
The CDR approach can be described in a very simple and direct way, just by showing that a varimax rotation of eigenvectors obtained from a ``generic" spectral algorithm works in some specific scenarios. 
However, we start this paper, and devote a large part of it, to providing a simple theory for {\em how} and {\em why} the CDR works. 
Aiming at a broader audience from different fields, and trying to use the most simple and transparent concepts as possible, this is done from a factor analysis point of view illustrated by a minimalistic toy-scenario. 
Both the theoretical aspect and the scenarios explored here do not represent a complete work, but the exploration for the feasibility of a CDR method to ``navigate" through differential structures in complex networks.

This paper is organized as follows. 
The mathematical theory, and illustrative motivational toy-scenario, for differential network analysis with the CDR are presented in Sec.~\ref{sec.methods}. Then, we explore the method with larger networks and in a larger number of conditions in Sec.\ref{sec.results}. 
Firstly, CDR is applied on synthetic networks with $N=2000$ nodes in $H=100$ different conditions, with a random number of communities (between 10 and 20) of random sizes (10 to 100 nodes) distributed in those conditions. 
The scope here is on networks of non-overlapping  clusters (or communities), generated from the stochastic block model. 
The clusters can be present or absent in different conditions. 
The method's accuracy is investigated through Monte Carlo simulations, where we manipulated the inner probabilities (how strongly the nodes within a community are connected) and the outer probability (how strongly nodes from different communities, and the ``noisy background", are connected). 
Then, 
we illustrate the strength of the CDR exploratory nature in revealing subtle differential structures of a gene expression dataset in four different conditions (two cell types $\times$ two treatments).  
Concluding remarks are made in Sec.\ref{sec.conc}.

\section{Methods}
\label{sec.methods}
In this section, we introduce the simple theory and basic methodology from a factor analysis (FA) point of view. 
The material is introduced by blending together a brief review of some FA results~\cite{koch_2013,Mulaik2010} with the pursuit of a theory for differential network analysis and community structure characterization. 
This means that we will start by being as abstract as necessary and that references to a ``toy scenario" (Sec.~\ref{sec.toy}) will be used as a motivation and to illustrate the main aspects of the proposed framework. 
Some small conceptual differences from actual factor analysis will be discussed when necessary.

\subsection{Motivation: toy scenario}
\label{sec.toy}
\begin{figure}[!htb]
	\centering
	\includegraphics[width=0.95\linewidth]{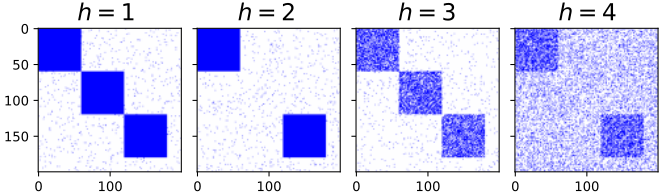}
	\caption{\label{fig.toy_scenario} Toy scenario of a small network in $H=4$ different conditions. It has $N=200$ nodes, which could be members of $R=3$ communities. This scenario can be thought of as a system in $H$ different experimental conditions, or an evolving network in $H$ different time points.}
\end{figure}

Consider the scenario of an undirected and unweighted network with $N=200$ nodes, in $H=4$ different conditions in respect to its connections, as shown in Fig.~\ref{fig.toy_scenario}. 
Assume that nodes neither disappear nor are created, only connections may change. 
For each condition $h = 1, ..., H$, the network is represented by its respective adjacency matrix ${\bf A}_h$ of size $N \times N$, with elements $a_{ij}=1$ if node $i$ and $j$ are connected, but $a_{ij}=0$ otherwise. 
Hence, the set $\{{\bf A}_h\}_{h=1}^H = \{{\bf A}_1, {\bf A}_2, {\bf A}_3, {\bf A}_4\}$ of adjacency matrices represents the structure of the network across the different $H$ conditions.  
There are $R=3$ highly connected and non-overlapping communities of sizes $N_r=60$ nodes each, as seen at Fig.~\ref{fig.toy_scenario}, labeled as $r=1,2,3$.

The set $\{{\bf A}_h\}_{h=1}^H$ was generated by a stochastic block model~\cite{HOLLAND1983109} with the package Networkx~\cite{SciPyProceedings_11}. 
The connection probabilities between nodes within the same community (inner probability) is $p$, being $q$ otherwise (i.e., the outer probability). 
Setting these probabilities as $0 \leq q < p \leq 1$ allows one to generate the $R$ highly connected communities, which can be visualized as as blocks in Fig.~\ref{fig.toy_scenario}. 
For the rest of the network structure, we considered a ``ghost" community of size $N-\sum_r N_r = 20$ nodes, with both the within and between probabilities equal to $q$. 
Let the adjacency matrices $\mathbfcal{A}_1$, $\mathbfcal{A}_2$ and $\mathbfcal{A}_3$ represent each of these communities. 
Accordingly, the scenario depicted here can be seen (conceptually) as $R$ Erd\"os and R\'enyi~\citep{Erdos:1960} random networks $G_{N_r,p}$ ``planted" on a $G_{N,q}$. 
Because of this, we will often refer to $G_{N_r,p}$ as the ``random background".

By design, the toy scenario illustrates two aspects regarding the structural changes in the network. 
The first one refers to the presence or absence of a community in a given condition. 
Specifically, communities $r=1$ and $r=3$ are present at all $H$ conditions, while $r=2$ is absent at conditions $h=2$ and $h=4$. 
In this aspect, the structure of the network is the {\it same} in (i) conditions 1 and 3, as well as in (ii) conditions 2 and 4, but (iii) different otherwise. 
That is the main structural change that we want to identify. 

However, the {\it internal} structure of conditions (and background noise) may be different, as follows.  
The second aspect refers to how strong a community is, as compared to the connections of its members to the other nodes (i.e., nodes from the other communities, as well as from nodes from the random background). 
This is done by using different combinations of the probabilities $p$ and $q$. 
The values for conditions 1 and 2 are $(p,q)=(1,0.02)$, therefore the communities are fully connected (known as 1-cliques). 
The communities became much more weakly connected in conditions 3 and 4 by setting $p=0.6$, while the background noise becomes stronger by a factor of 10 at condition 4 with $q=0.2$. 
So, the {\it mixing} between communities and background noise is larger at condition 3, and much larger at condition 4.  
Later we will investigate how the mixing level interferes on the CDR method. 

\begin{remark}
We call attention to one consequence of the SBM, that could otherwise pass unnoticed. For example, the {\it internal structure} of the community $r=1$ can be completely different between conditions 3 and 4. 
They just share the same inner probability $p=0.6$, but the actual connections of their nodes are set at random by the model. 
The same occurs for the random background in conditions 1, and 3: the inner and outer probabilities $p=q=0.02$ are the same, but the actual connections are not. 
\end{remark}

\subsection{Theoretical framework}
The initial assumptions of the method are as follows. 
There exists an abstract {\em property} $y_i$, $i = 1, ..., HN$, for the $N$ nodes in the $H$ {\em conditions}. 
For example, this means that for a given node $j \in [1, N]$, the properties $y_{j+(h_1-1)N}$ and $y_{j+(h_2-1)N}$ will refer to the same node $j$ in two different conditions $h_1 \neq h_2 \in [1, H]$. 
Conceptually, we will assume that a causal relationship exists between the set $\{y_i\}_{i=1}^{HN}$ and the sets of unknown and abstract processes $\{\xi_i\}_{i=1}^v$ and $\{e\}_{i=1}^{HN}$ (i.e., the factors or latent variables). 
For now, we just assume that $y$, $\xi$ and $e$ can be represented as vectors in an abstract vector space, mainly because we will need the concept of inner products $\left\langle \bullet |\bullet \right\rangle$ to represent the extent to which they are (or they are not) close. 
In the set $\{e_i\}$ are the {\em factors} specific for each $y_i$ (i.e., the {\it unique factors}). 
The set $\{\xi_i\}$ are factors that can influence any and several $y_i$ (i.e., the {\it common factors}). 
We assume that the common and unique factors are independent, $\left\langle \xi_i |e_j \right\rangle = 0$ , and that the unique factors are orthogonal, $\left\langle e_i |e_j \right\rangle =0$ if $i\neq j$ 
(orthogonality of the common factors will not be assumed yet). 
By defining the column matrices $\mathcal{Y} = [y_1 ...\, y_{HN}]^{\top}$, $\mathcal{X} = [\xi_1 ...\, \xi_v]^{\top}$ and $\mathcal{E} = [e_1 ...\, e_{HN}]^{\top}$, one can write the linear model

\begin{equation}
\label{eq.full-factor-model}
\mathcal{Y}=\mathbf{\Lambda} \mathcal{X}+\mathbf{\Psi} \mathcal{E},
\end{equation}
known as the {\it fundamental equation} of factor analysis~\cite{Mulaik2010}. 
The matrices $\mathbf{\Lambda}$ (size $N\times v$) and $\mathbf{\Psi}$ (size $HN \times HN$) provide the common and unique factor loadings (or weights), respectively. 
Matrix $\mathbf{\Psi}$ is diagonal (i.e., off-diagonal elements are equal to zero), because the factors $e_i$ are unique. 
Without loss of generality, we assume the factors' norm are equal to one, because they can be absorbed by the factor weight matrices $\mathbf{\Lambda} $ and $\mathbf{\Psi} $. 
So, $\left\langle e_i |e_j \right\rangle =\delta_{i,j}$ and $\left\langle \xi_i | \xi_i \right\rangle =1$ ($\delta$ is the Kronecker's delta function). 

There is a subtle conceptual different between  model (\ref{eq.full-factor-model}) and an actual factor analytical model. 
In the latter, $y$ refers to {\it measured data}. 
But in this paper, we will consider $y$ just as an {\it abstract concept} parameterised in a vector space, which will provide us with flexibility latter. 
Accordingly, no assumptions will be made regarding the particular statistics of $y$ (e.g., a random variable with zero mean and unity variance. 
As well, we will use the concept of a Gramian matrix, instead of the covariance or the correlation matrices, for the matrices related to the inner products. 
Actual measured data will be inserted into the framework in the next section.

Finally, we assume that the goal of modeling a given phenomenon (with data 
from a set of observations) through network theory is to investigate the {\it differential} clustering of nodes: 
what structures remain the same, and what changes, along with the different conditions $H$. 
Now we explore two consequences of model (\ref{eq.full-factor-model}) under those assumptions.

\subsubsection{Feasibility for differential network analysis and clustering}
The characterization of the clustering of nodes along conditions due to the sharing of latent variables $\xi_i$ could be achieved by inspecting the structure of the matrix $\mathbf{\Lambda}$. 
To see this, consider the product between $y_i$ and $\xi_j$:  
$\mathbf{R}_\mathcal{YX} \doteq\mathcal{Y}\mathcal{X}^{\top}$, with elements $[\mathbf{R}_\mathcal{YX}]_{i,j} = \left\langle \xi_i |y_j \right\rangle$. 
Because $\left\langle \xi_i |e_j \right\rangle = 0$, we have


\begin{equation}
\label{eq.Ryx1}
\mathbf{R}_\mathcal{YX}\doteq\left[\begin{array}{ccc}
\left\langle \xi_1 |y_1 \right\rangle &       \cdots & \left\langle \xi_v |y_1 \right\rangle \\
\vdots &  \ddots & \vdots \\
\left\langle \xi_1 |y_{HN} \right\rangle &    \cdots & \left\langle \xi_v |y_{HN} \right\rangle
\end{array}\right]
\equiv \mathbf{\Lambda} \mathbf{R}_\mathcal{XX},
\end{equation}
where $\mathbf{R}_\mathcal{XX} \doteq\mathcal{X}\mathcal{X}^{\top}$.

Expression (\ref{eq.Ryx1}) can be simplified even further if we are allowed to assume orthogonality between the common factors, $\left\langle \xi_i |\xi_j \right\rangle = \delta_{i,j}$. 
Under that new assumption, for which henceforth we restrict the scope of this paper, (\ref{eq.Ryx1}) becomes 

\begin{equation}
\label{eq.Ryx2}
\mathbf{R}_\mathcal{YX}=\mathbf{\Lambda} 
\end{equation}

\begin{figure}
	\centering
	\includegraphics[width=0.85\linewidth]{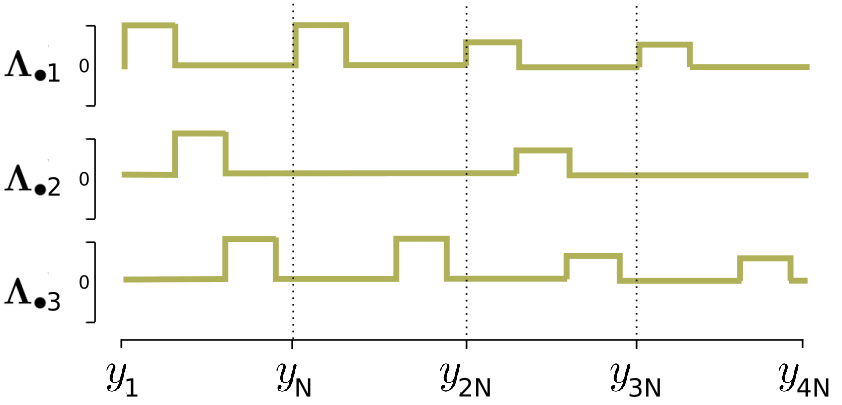}
	\caption{\label{fig.insigths-Ryx} Schematic picture of the intuitive expected structure of matrix $\mathbf{\Lambda}$ in the context of the toy scenario of Fig.~\ref{fig.toy_scenario}. 
	The leading three columns $\mathbf{\Lambda}_{\bullet k} = [\left\langle \xi_k | y_1 \right\rangle  ... \left\langle \xi_k | y_{HN} \right\rangle]^{\top}$, with $k=1,2,3$, are shown.}
\end{figure}

To gather insights of the implications of (\ref{eq.Ryx1}) and (\ref{eq.Ryx2}), we use (\ref{eq.full-factor-model}) to frame the problem of differential network analysis and community characterization of the toy scenario shown in  Fig.~\ref{fig.toy_scenario}.  
In that context, and because we now know the real community structures across conditions (i.e., the ground-truth), a reasonable hypothesis is that the ``true" underlying community structure  to be captured by (\ref{eq.full-factor-model}) is given by the $R$ known planted communities only, and not by the background noise from $G_{Nq}$. 
Then, by-design the underlying theoretical assumptions for the FA model (\ref{eq.full-factor-model}) would be: 
(i) $\xi_1$ is the ``cause" of the community structure of nodes 1 to 60 (i.e., $r=1$) at all $H=4$ conditions; 
(ii) $\xi_2$ for nodes 61 to 120, but {\it only} at conditions $h=1,2$. 
(iii) $\xi_3$ for nodes 121 to 180 at all $H=4$ conditions. 
Note that the labels 1, 2 and 3 were used here for the sake of illustration (e.g., community $r=1$ could  be related to $\xi_3$ and so on). 
Because of that, expressions (\ref{eq.Ryx1}) and (\ref{eq.Ryx2}) tell us that these causal relationships should be reflected on the structure of the common factor loading matrix $\mathbf{\Lambda}$ as high loadings related to those three factors $\xi$ for the properties $y_i$ within these ranges. 
That is schematically shown in Fig.~\ref{fig.insigths-Ryx}, where $\mathbf{\Lambda}_{\bullet k} = [\left\langle \xi_1k |y_1 \right\rangle  ... \left\langle \xi_k |y_{HN} \right\rangle]^{\top}$.

The main result of this paper is based upon finding (or extracting) {\it that} special structure from measured data, because it clearly reports on the structural changes of the network. 
In a sense, this provide us with a map, or a book composed by $H$ ``chapters" within the $H$ segments of length $N$, that allows one to navigate through the network differential structure history.  
Therefore, our aim now is to obtain that idealized structure after fitting {\it observed data} on a model based on (\ref{eq.full-factor-model}), and then both (i) community structure characterization and (ii) differential network analysis would be straightforward.

There is a subtle conceptual different between  model (\ref{eq.full-factor-model}) and an actual factor analytical model. 
In the latter, $y$ refers to {\it measured data}. 
However, in this paper we are considering $y$ just as an {\it abstract concept} within a vector space, which will provide us with flexibility latter. 
Accordingly, no assumptions will be made regarding particular statistics of $y$ (e.g., a random variable with zero mean and unity variance). 
As well, we will make use of the concept of a Gramian matrix, instead of the covariance or the correlation matrices, for the matrices related to the inner products. 
Actual measured data will be inserted into the framework in the next section, when we show one way to extract  the structure $\mathbf{\Lambda}$.

\subsubsection{Extracting the structure of $\mathbf{\Lambda}$ }
Consider the projection of $y_i$ onto itself, $\mathbf{R}_\mathcal{YY} \doteq \mathcal{Y}\mathcal{Y}^{\top}$ (with matrix elements $[\mathbf{R}_\mathcal{YY}]_{i,j} = \left\langle y_i |y_j \right\rangle$). 
Because the common and unique factors are orthogonal one can write $\mathbf{R}_\mathcal{YY}=\mathbf{\Lambda}^2 \mathbf{R}_\mathcal{XX}+\mathbf{\Psi}^2$, an expression that is often known as the {\it fundamental theorem} of factor analysis~\cite{Mulaik2010}. 
Rearranging for $\mathbf{\Lambda}^2$, and because one previously assumed the orthonormality between the common factors, we have

\begin{equation}
\label{eq.Ryy_psi}
\mathbf{\Lambda}^2 = \mathbf{R}_\mathcal{YY} -\mathbf{\Psi}^2.
\end{equation}
We will simplify (\ref{eq.Ryy}) even further by assuming that the term $\mathbf{\Psi}^2$ is neglegible, so 
\begin{equation}
\label{eq.Ryy}
\mathbf{\Lambda}^2 = \mathbf{R}_\mathcal{YY}.
\end{equation}
That is an assumption very often used in FA.  
If there exists a way to estimate the contributions from the unique factors, one can go back and simply update the diagonal elements of $\mathbf{\Lambda}^2$ (because $\mathbf{\Psi}^2$ is a diagonal matrix). 
Finally, writing the eigendecomposition $\mathbf{R}_\mathcal{YY} = \mathbf{V}\mathbf{\Sigma}\mathbf{V}^{\top}$, we have
\begin{equation}
\label{eq.lambda-is-SigmaE}
\mathbf{\Lambda} = \mathbf{\Sigma}^{1/2} \mathbf{V}.
\end{equation}
The eigenvectors $\mathbf{v}_k$ correspond to the columns of matrix $\mathbf{V}$, with respective eigenvalues $\sigma_k$, $k=1,2, ...n$ in the main diagonal of matrix $ \mathbf{\Sigma}$. 
We assume they are in the decreasing order $\sigma_1>\sigma_2 ... >\sigma_n$.

That is one of the several procedures for extracting the structure of the factor loading matrix. 
It is sometimes called {\it principal component factor analysis}~\cite{koch_2013} --- 
or referred to as extraction through {\it principal components analysis} (PCA)~\cite{Mulaik2010}.  
However, that is the case when $\mathbf{R}_\mathcal{YY}$ comes from {\it measured data} (equivalently, $y_i$ is a random vector). 
So, now we need to address the aforementioned conceptual different between model (\ref{eq.full-factor-model}) and an actual factor analytical model: $y_i$ are concepts, not random variables.

{\it What does it mean that two nodes belong to the same community?} 
The answer can (and should) depend on the actual research question and the field-dependent characteristics that one aims at by grouping the nodes and asking for their differential network structure. 
Here, letting $y_i$ being concepts and not actual data, we aim at that flexibility for the CDR framework to address different field-specific points-of-view. 
This can be put more clearly through the following example, which will be used as well to establish remaining procedures.

\begin{figure}
	\centering
	\includegraphics[width=0.85\linewidth]{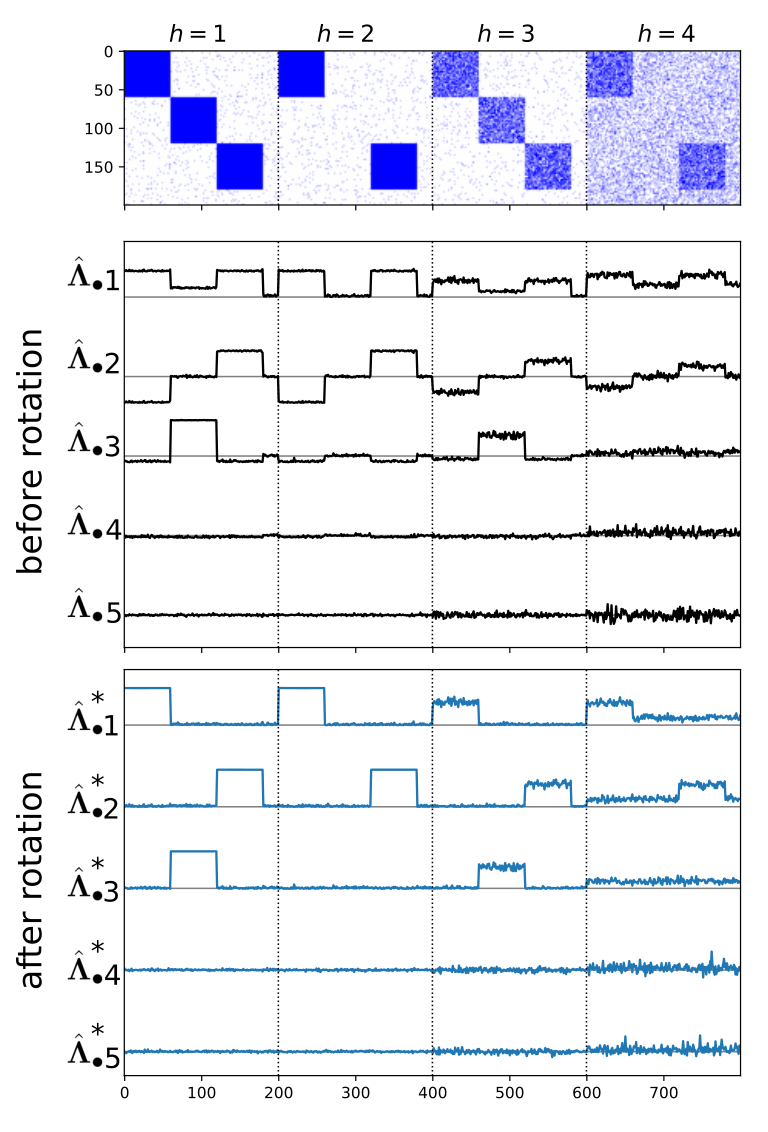}
	\caption{\label{fig.toy-res-1} Results of the CDR on the toy scenario of Fig.~\ref{fig.toy_scenario}. 
	The concatenated matrix $\mathbf{X}$ is shown on top. 
	The structure of matrix $\mathbf{\Lambda}$ is shown before (middle) and after (bottom) the varimax rotation. 
	Before rotation, the leading loadings $\hat{\mathbf{\Lambda}}_{\bullet k}$ contain mixed information regarding the $R=3$ communities. 
	After the varimax rotation (bottom row), the rotated loadings $\hat{\mathbf{\Lambda}}^*_{\bullet k}$ are unmixed and bring more detailed information about the communities and on how they change between the $H=3$ conditions --- 
	a structure that clearly mirrors the idealised one of Fig.~\ref{fig.insigths-Ryx}. 
	}
\end{figure}

Consider again the toy scenario of Fig.~\ref{fig.toy_scenario}, and assume that the measured data is the set  $\{{\bf A}_h\}_{h=1}^H$. 
For the task at hand (differential network analysis) and the characteristics of the community structures (high within connected blocks, low between connectivity), one reasonable choice is to use the similarity of the list of neighbours between nodes and conditions to capture the relevant question one wants to address through a differential network analysis. 
Given two nodes $i$ and $j$, their list of neighbours in conditions $h_1$ and $h_2$ are the columns $[\mathbf{A}_{h_{1}}]_{\bullet i}$ and $[\mathbf{A}_{h_{2}}]_{\bullet j}$ of the respective adjacency matrices. 
Let $\mathbf{X} = [\mathbf{A}_1~\mathbf{A}_2~...~\mathbf{A}_H]$ be the $N\times HN$ matrix formed by horizontally concatenating the $H$ adjacency matrices (see Fig.~\ref{fig.toy-res-1}, top panel).  
Accordingly, one defines the estimate $\mathbf{\hat R}_\mathcal{YY}$ for $\mathbf{R}_\mathcal{YY}$ as 

\begin{equation}
\label{eq.Ryy_hat}
\mathbf{R}_\mathcal{YY} \doteq  \mathcal{Y}\mathcal{Y}^{\top} \mathrel{\hat=} \mathbf{\hat R}_\mathcal{YY} \triangleq \mathbf{X}^\top\mathbf{X},
\end{equation}
where the symbol $\mathrel{\hat=}$ stands for ``estimated from". 
Here, we are using the symbol $\triangleq$ to emphasize that this definition depends on the field-specific characteristics that are pertinent for the question one wants to answer.  
Henceforth in this paper, we will make use of (\ref{eq.Ryy_hat}) to estimate the common factor loadings ${\hat{\mathbf{\Lambda}}}$ through the eigendecomposition of $\mathbf{X}^\top\mathbf{X}$.

The leading five columns of the estimated loadings,  $\hat{\mathbf{\Lambda}}_{\bullet k}$ for  $k=1, ..., 5$,   are shown in Fig~\ref{fig.toy-res-1} (middle row). 
Henceforth we will call them simply  ``loadings". 
They contrast deeply with the desired ``simple" structure previously shown in Fig.~\ref{fig.insigths-Ryx}. 
Specifically, one sees a mixed signature of communities, which is more entangled in the leading two loadings. 
Actually, that is indeed the expected {\it intermediate} result from FA. 
The reason is that we have applied PCA to extract the loadings and PCA, by itself, is the solution for the maximization problem $\max \operatorname{tr}cov(X)$ given the restriction of orthonormality of the principal directions $\left\langle \mathbf{v}_i |\mathbf{v}_j \right\rangle =\delta_{i,j}$. 
So, PCA maximizes the variance explained by the leading $k$ components, and a large amount of the information become entangled in the leading eigenvectors $\mathbf{v}_i$ --- and,
consequently, in the leading $\hat{\mathbf{\Lambda}}_{\bullet k}$.

The usual follow up procedure in FA is based on Thurstone's concept of simple structure~\cite{Thurstone1931}: the rotation of the factor loadings by a given criterion that maximizes the {\it simplicity} of $\hat{\mathbf{\Lambda}}$, and so (hopefully) enhancing the interpretability of that matrix.

\subsubsection{Varimax rotation and the simple structure of $\mathbf{\Lambda}$}
Kaiser’s varimax~\cite{Kaiser1974} is considered the most-efficient (orthogonal) rotation in FA~\cite{Richman1986}, and the most often applied. 
Let the elements of the factor loading matrix be $\hat{\mathbf{\Lambda}} = [\lambda_{k,d}]$. 
The varimax rotation aims at finding the orthogonal rotation $\hat{\mathbf{\Lambda}}^* = \hat{\mathbf{\Lambda}}\mathbf{T}$ that satisfies the varimax criterion (VC)

\begin{equation}
\label{eq.varimax}
{\rm VC}(\mathbf{\Lambda})=\sum_{k=1}^{S}\left[\frac{1}{D} \sum_{d=1}^{D} \lambda_{dk}^{4}-\left(\frac{1}{D} \sum_{d=1}^{D}  \lambda_{dk}^{2}\right)^{2}\right]. 
\end{equation}
Specifically, (\ref{eq.varimax}) is the {\it raw} varimax criterion. 
It represents the maximization of the variance across the columns of the squared factor loadings matrix. 
The summation is over the first $S$ factors $\hat{\mathbf{\Lambda}}_{\bullet k}$, $k=1,...,S$.

The result of that rotation is shown in Fig.~\ref{fig.toy-res-1}(bottom), with $S=20$. 
Each leading $\hat{\mathbf{\Lambda}}^*_{\bullet k}$ carries now a unique fingerprint of (i) each  community for (ii) each condition, similar to the expected idealized structure in Fig.~\ref{fig.insigths-Ryx}. 
The main result of this paper is that that approach recovers the full ``story" of the network, where each ``chapter" is encoded on each $H=3$ segment of length $N$. 
Then, it becomes straightforward to read: communities 1 and 2 were present along with all $H=3$ conditions, while community 2 disappeared in condition $h=2$ but reappeared in $h=3$.  
Another information is provided by the different magnitude of the pulse-like pumps at different segments of the same $\hat{\mathbf{\Lambda}}^*_{\bullet k}$. 
For example, consider the $\hat{\mathbf{\Lambda}}^*_{\bullet 1}$. 
The first two pumps have the same magnitude, which is larger than the magnitude of the last two segments. 
This means that, besides the community $r=1$ been present in all $H=4$ conditions, something in its structure is more similar within conditions $h\in\{1,2\}$ and $h\in\{3,4\}$ than between them. 
By design, we know that this should be a consequence of the different inner probabilities: $p=1$ for $h\in\{1,2\}$, and $p=0.6$ for $h\in\{3,4\}$. 
As a remark, note that as because of the stochastic block model applied, the structure of a given community is not identical in different conditions (when $p < 1$, i.e., communities are {\it not} 1-cliques), neither the connections of its nodes with the outer nodes (for any value of $p$).  
Even so, the proposed approach shows a clear representation of the differential community structure.

There is another feature in Fig.~\ref{fig.toy-res-1} worth of commenting. 
A constant trend (vertical shift) in the leading loadings (both before and after rotation) is clearly seen when the larger outer probability $q=0.2$ is used to build the network. 
That trend is a known consequence of not using a column-centralized matrix in PCA. 
Actually, the trend is present in the other loadings too. 
It is more visible for larger values of $q$ because larger the $q$, larger the mean value of each column of $\mathbf{X}$ (i.e., more connections imply at more ``ones" instead of ``zeros"). 
For the rotated loadings, it is clearly visible in the fourth (last) segment of length $N$ of the leading 3 rotated loadings $\hat{\mathbf{\Lambda}}^*_{\bullet k}$. 
For our goal in this paper, that trend is irrelevant. 

\begin{remark}
The analysis here could be conducted by the point-of-view of the eigenvectors $\mathbf{v}_k$ (the columns of $\mathbf{V}$). 
In that case, the correct way to obtain the varimax rotated $\mathbf{v}_k^*$ is using the scaled $\sigma^{\frac{1}{2}}\mathbf{v}_k$ to find the rotation $\mathbf{T}$, and then applying the rotation to the original (not scaled) eigenvectors, $\mathbf{v}_k^* = \mathbf{v}_k\mathbf{T}$. 
The pitfall here is that, because the scaled eigenvectors correspond to the common factor loadings, one could assume {\em incorrectly} that the rotated eigenvectors could be obtained simply by rescaling the already rotated factor loadings. 
However, this process of scaling, rotating and rescaling yields an {\em oblique} rotation. 
\end{remark}

\subsection{Statistical analysis}
\label{sec.res.stat}
Numerically generated scenarios with much larger $R$, $N$ and $H$ will be used in the next section to investigate the application of the proposed CDR framework. 
For some of them, we will make use of large Monte Carlo runs, using the outer probability $q$ as the mixing parameter. 
While the emphasis in this paper is on the interpretability provided by the visual inspection of the rotated  $\hat{\mathbf{\Lambda}}^*_{\bullet k}$  as an exploratory tool (as in Fig.~\ref{fig.toy-res-1}, bottom), those larger scenarios bring the need of using auxiliary tools to validate the method and to automate some steps. 

Firstly, in order to quantify the accuracy of the rotated $\hat{\mathbf{\Lambda}}^*_{\bullet k}$  on correctly identifying the communities $r$ that share the same condition $h$, we use the score provided by the area under the receiver operating characteristics curve (AUC-ROC). 
The ROC curve is a common tool in machine learning, used to compare the performance of binary classifiers. 
It is a graphical representation where the true positive rate (also called sensitivity or recall) is plotted against the false positive rate for different thresholds used in the decision function. 
That score can have values between 1 and 0.5: a random classifier will have AUC-ROC equal to 0.5, whereas a perfect classifier will have ROC-AUC equal to~1. 

\begin{figure}
	\centering
	\includegraphics[width=0.85\linewidth]{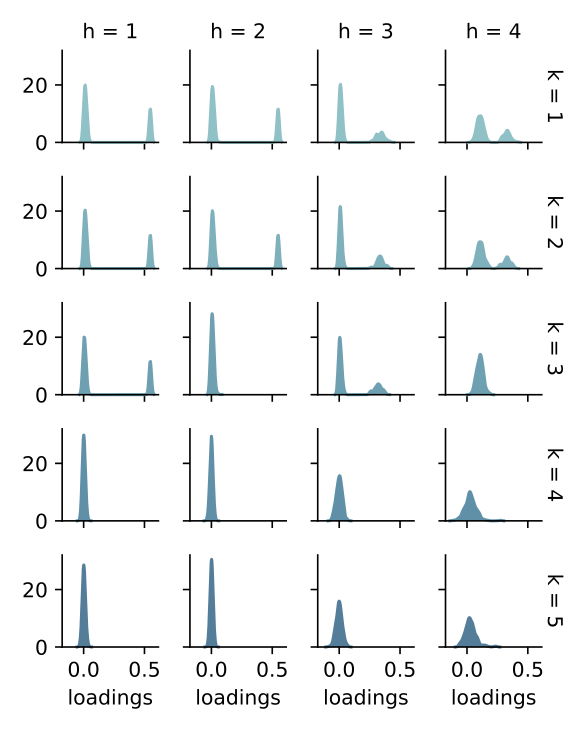}
	\caption{\label{fig.toy_kde} Estimated gausian kernel densities for each segment $h$ for the leading 5 rotated factor loadings $\hat{\mathbf{\Lambda}}^*_{\bullet k}$, corresponding to Fig.~\ref{fig.toy-res-1} (botom). 
	Darker shades refers to larger values of $k$. 
	}
\end{figure}

Let  $\hat{\mathbf{\Lambda}}^*_{\bullet k,h}$  denote each segment of length $N$ of the rotated factor loadings matrix $k^{\rm th}$-column, with elements $\hat{\mathbf{\Lambda}}^*_{\bullet k,h}(i)$ , $i=1,2, ..., N$. 
Those segments are binary classifiers, which can be used to cluster the nodes into one or two groups given their respective weights $\hat{\mathbf{\Lambda}}^*_{\bullet k,h}(i)$ 
(by referring to that clusters as ``groups" we aim to avoid any confusion with the planted communities $r$).  
For instance, consider the segment $\hat{\mathbf{\Lambda}}^*_{\bullet 1,1}$ in Fig.~\ref{fig.toy-res-1} (bottom). 
The weights $\hat{\mathbf{\Lambda}}^*_{\bullet 1,1}(i)$ clearly form two clusters: one with values near the ``noise floor", and the other with much larger values than the noise floor variance. 
Figure~\ref{fig.toy_kde} shows the distribution of weights for all segments in Fig.~\ref{fig.toy-res-1} (bottom). 
Let's denote those two groups seen in $\hat{\mathbf{\Lambda}}^*_{\bullet 1,1}(i)$ by labels 0 and 1. 
We define the index vector $\mathbf{s}^{(k,h)}$ of length $N$ with elements

\begin{equation}
\label{eq.s-vector}
  s_i^{(k,h)} =
  \begin{cases}
    0, & \text{if node } i \text{ belongs to group } 0 \\
    1, & \text{otherwise.}
  \end{cases}
\end{equation}
By the other hand, note that the ``flat" segment $\hat{\mathbf{\Lambda}}^*_{\bullet 3,2}$, which is the fingerprint of the absence of $r=2$ in condition $h=2$, will generate a $\mathbf{s}^{(3,2)}$ with all its elements equal to 0 (i.e., all nodes are associated to the noise floor, and so to group 0). 
The differential structure in Fig.~\ref{fig.toy-res-1} is so clearly discernible that the clustering could be done by visual inspection. 
However, for large numbers of conditions and communities, this would be prohibitive. 
An alternative solution would be applying a given clustering algorithm of choice.

Hence, secondly, in this paper, we opt to use the density-based spatial clustering of applications with noise~\cite{schubert2017dbscan,ester1996density} (DBSCAN), with parameters {\it number of neighbours} and {\it distance} equal to 5 and $1/100$ of the amplitude of the $\hat{\mathbf{\Lambda}}^*_{\bullet k,h}$, respectively. 
By design in the next section numerical experiments, the ground truth is known, and hence it is used to construct the index vectors $\mathbf{s}_{\rm true}^r$ for each $r = 1, 2, ...R$ planted community. 
Then, we compute the AUC-ROC score between a single $\mathbf{s}^{(k,h)}$ and all the $R$ others $\mathbf{s}_{\rm true}^r$ vectors. 
We pick the largest value and the value of $r$ for which it occurred, and use them to represent both the method score in identifying the community with label $r$ through the the $\hat{\mathbf{\Lambda}}^*_{\bullet k,h}$ segment. 
Note that the knowledge about the condition $h$, where that community $r$ was found, is already provided by the $h$ index of the current $\xi_{k,h}^*$. 

Finally, the overall accuracy will be quantified by the fraction of {\it reliably} detected communities. 
Specifically, we count the number of detected communities with a ROC-AUC score above a given threshold and divide that value by the real (known) number of communities. 
The value of this threshold is $0.8$.

\section{Results}
\label{sec.results}
The motivation in this section is two-fold. 
First, numerical experiments are conducted to test the accuracy of the CDR method in unveiling the history of structural changes in more challenge scenarios. 
This is done for a larger network of $2000$ nodes in $H=100$ conditions that could have between 10 and 20 communities with very different sizes. 
Whilst that is a much more complex scenario than the previous toy one, it is only a small representation of the myriad ways in which the network structure could change in real-world systems. 
For instance, in this paper, we are focusing on communities that could only be present or absent.  In reality, there can be superpositions, growing and shrinking, combinations of those process etc.  
Notwithstanding, as the motivation of this paper is to provide a first step for the CDR method, we make our best to deeply explore this constrained scenario.

The second motivation is to explore the potential use of the navigation map provided by the CDR approach, in the context of gene coexpression data. 
The dataset used consists of expression data of 1458 genes, from 2 different cell types and in 2 different experimental conditions. 
This application is our principle motivation for the CDR. 
In our specific setting, this means a network with 1458 nodes in $H=4$ conditions. 
Whilst the adjacency matrices in that context are weighted rather than binary, and as well there could be both positive and negative connections, we show that the CDR can find unique and subtle nuances of the structural differences. 
As a remark, this is done here for illustrative purpose, and not to unveil any biological meaningful result for the specific dataset we used.

\subsection{Synthetic networks}
\label{sec.synthetic}
\begin{figure*}
	\centering
	\includegraphics[width=0.85\linewidth]{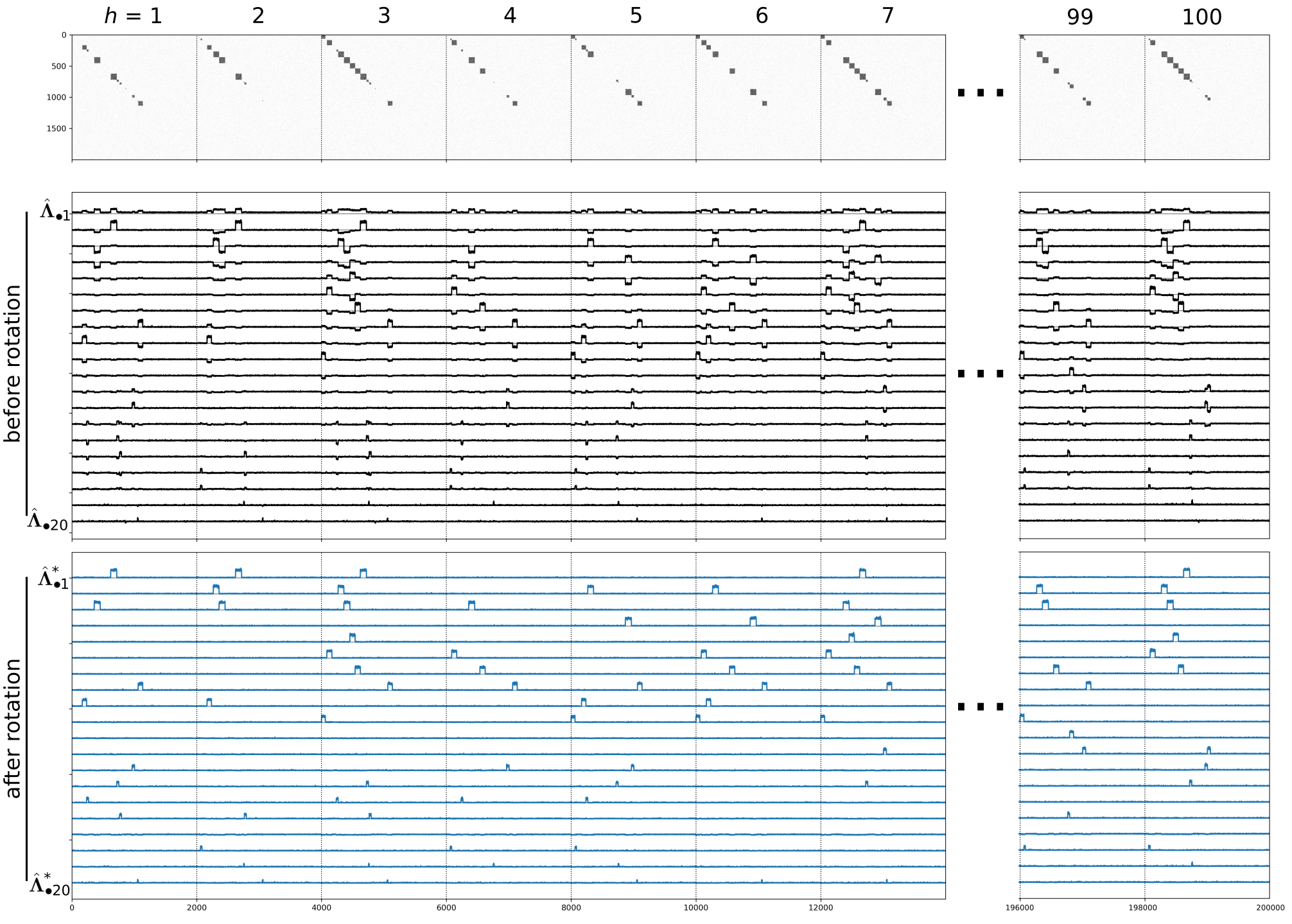}
	\caption{\label{fig.100cases-factors} Concatenated matrix $\mathbf{X}$ (top) for the large network ($N=2000$) with $R=20$ communities randomly distributed along $H=100$ conditions. 
	Communities sizes $10\leq N_r \leq 100$ were sampled from an uniform distribution. 
	Inner and outer probabilities are $(p, q) = (0.6, 0.02)$. 
	Before rotation (middle) the factor loadings $\hat{\mathbf{\Lambda}}_{\bullet k}$ show mixed signatures of the communities. 
	After rotation (bottom), each community in each ``shared" condition is clearly represented within the same $\hat{\mathbf{\Lambda}}^*_{\bullet k}$.  }
\end{figure*}

We start by mimicking the scenario of a large network with $N=2000$ nodes in $H=100$ conditions. 
In a similar fashion as in the toy scenario of Sec.~\ref{fig.toy_scenario}, a set of 100 adjacency matrices $\{{\bf A}_h\}_{h=1}^{100}$ was generated using the SBM.
However, the difference is that the number of planted communities in each condition, the community sizes and their specific labels were chosen from random uniform distributions. 
Figure~\ref{fig.100cases-factors} (top) shows the 10 blocks of the concatenated matrix  $\mathbf{X} = [\mathbf{A}_1~\mathbf{A}_2~...~\mathbf{A}_{100}]$ associated with the first 7 and last 2 adjacency matrices. 
The inner and outer probabilities are $p=0.6$ and $q=0.02$.

The specific steps to generate this scenario are:
\begin{enumerate}
\item An array containing 20 tuples $(r, N_r)$ was generate with $N_r \sim \mathcal{U}(10,100)$, associating a given community label $r=1, ..., 20$ with its respective community size. 

\item An array containing 100 tuples $(h, R_h)$, with $R_h \sim \mathcal{U}(10,20)$, specifies the number of communities $R_h$ to be planted in a condition $h$. 

\item Fixing $h$, the adjacency matrix ${\bf A}_h$ was generated with $R_h$ communities as specified by the step (2), but with community, labels randomly sampled (with equal probability) from the array generated in step (1). 

\item Step (3) was repeated for $h=1, ..., 100$.
\end{enumerate}

Note that, as in the toy scenario, the nodes belonging to the same community with label $r$, but in different conditions $h$, will have a different specific internal (and external) connections. 
Only their inner and outer probabilities are the same across conditions. 
 
In summary, each condition $h=1, ..., H$ can have 10 to 20 planted communities, and each of them can have (or not) different sizes. 
A given community can appear in several conditions, but its internal and external connectivity will be (very likely) different.

The leading 20 loadings before and after rotation are shown in Fig.~\ref{fig.100cases-factors} (middle and bottom, respectively). 
The varimax rotation was performed with $S=2R=40$ vectors, assuming that the double of the maximum allowed number of communities (in any condition) will provide a sufficient degree of freedom for the algorithm to achieve the desired simple structure. 
The expected mixing before rotation, and clear representation of the differential network structure after rotation, can be seen. 
Worth mentioning, the fingerprints of the small communities at the last loadings $\hat{\mathbf{\Lambda}}^*_{\bullet k}$ (e.g., for $k=20$) would {\it not} be visually discernible from the random ``noise floor" in a real application: 
here, by design, successive node indexes were associated with the communities. 
In contrast, with real-world data, the indexes of nodes from the same community will be (very likely) spread along the integer segment $[1,N]$.

\begin{figure*}
	\centering
	\includegraphics[width=0.85\linewidth]{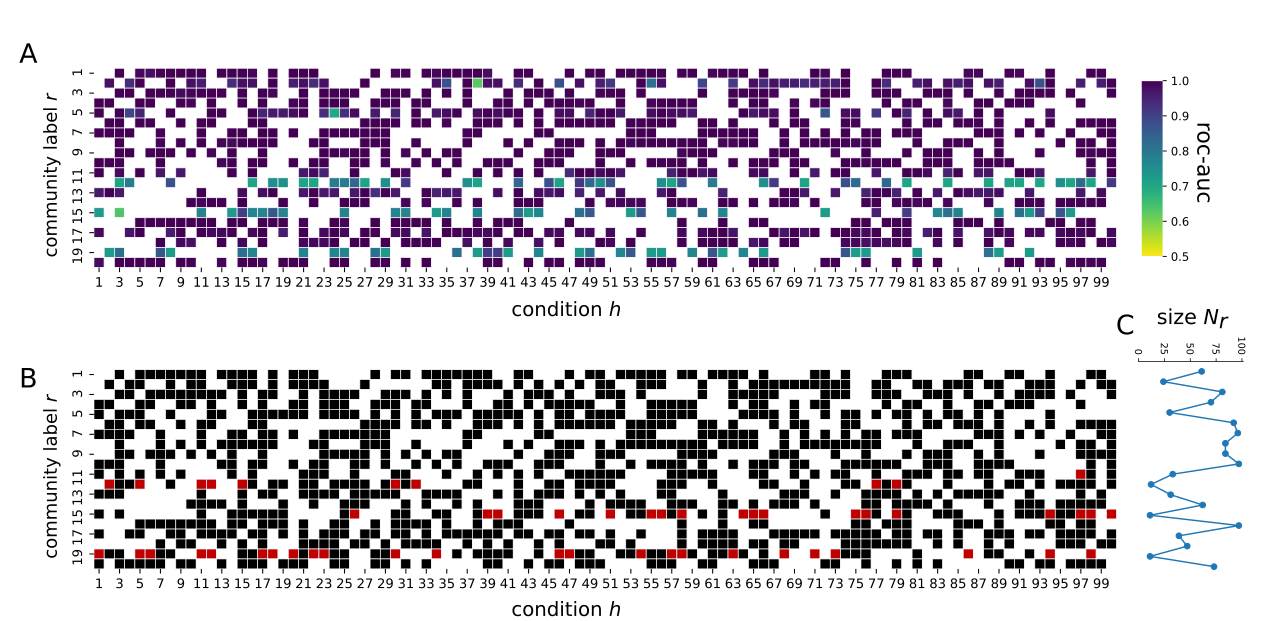}
	\caption{\label{fig.100cases-roc} Communities planted and detected for the large network in $H=100$ conditions (see Fig.~\ref{fig.100cases-factors}). 
	(A) Roc-auc scores of the communities detected by the rotated loadings $\hat{\mathbf{\Lambda}}^*_{\bullet k}$. For clarity, only scores above 0.6 are shown.  
	We considered a {\it successful} detection if the ROC-AUC score is above 0.8. 
	(B) The ground truth: communities planted in each condition $h$. 
	Colors indicate if they were successfully detected (black) or not (red). 
	(C) The community sizes $N_r \sim \mathcal{U}(10,100)$ used in this simulation. 
	Smaller communities are more likely to be missed by the method. 
	}
\end{figure*}

   
Now we assess the fidelity of this representation provided by the rotated $\hat{\mathbf{\Lambda}}^*_{\bullet k}$. 
The AUC-ROC scores are shown in Fig~\ref{fig.100cases-roc}.A (for the sake of clarity, only values above $0.6$ are shown). 
The distribution of values will be discussed later. 
Still, one can see that the majority of them are near 1: the best achievable balance between almost perfect (i) sensitivity (100\% true positive rate) and specificity (0\% false positive rate). 
Those values can be contrasted with the ground-truth shown in Fig.~\ref{fig.100cases-roc}.B. 
The filled squares indicate the planted communities: black if the community was identified by the method (i.e., a ROC-AUC score above 0.8), and red otherwise. 
The fraction between detected and planted communities (the number of black squares divided by the total number of squares) is 0.84. 
That means 921 successful identifications, and 130 misses. 
By comparison with the actual community sizes $N_r$, given by plotting the array with 20 tuples $(r, N_r)$ in Fig~\ref{fig.100cases-roc}.(C), we see that the small communities, with size near 10, are more likely to be missed.

\begin{figure}[!htb]
	\centering
	\includegraphics[width=0.85\linewidth]{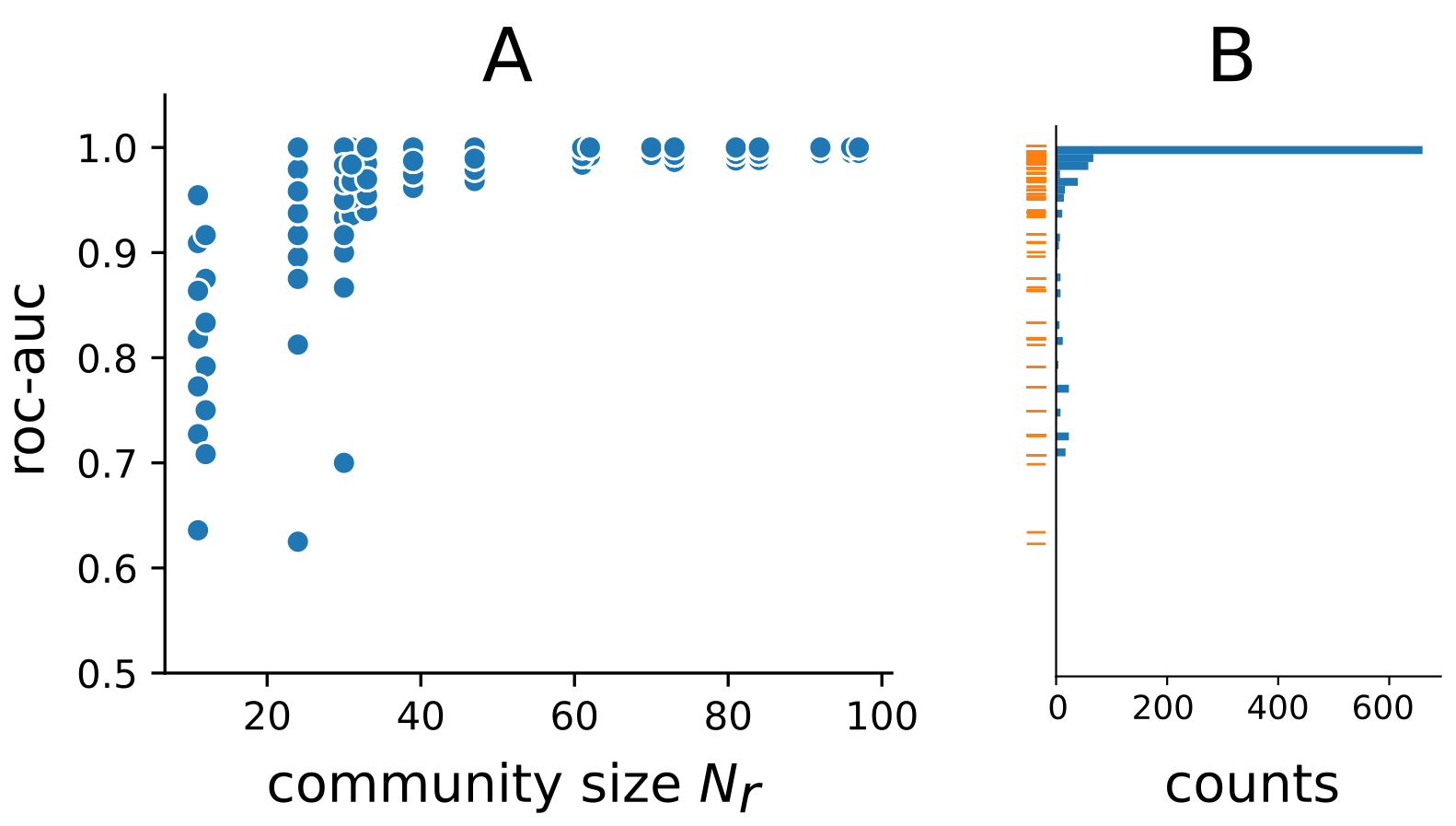}
	\caption{\label{100cases-roc-vs-Nr} Distribution of ROC-AUC scores shown in Fig~\ref{fig.100cases-roc}.B. 
	The fidelity of community detection (A) is higher for larger communities. 
	 Because several scores an equal to 1, the respective makers are overlapped, and in (B) one can see better their distribution. 
	 Remark: for the sake of clarity, only values with ROC-AUC larger than 0.6 are plotted. 
	  }
\end{figure}

Figure~\ref{100cases-roc-vs-Nr} shows the distribution of ROC-AUC scores relative to the community sizes. 
Two features can be seen, which confirms the previous discussion. 
Firstly, communities with size above 60 nodes were detected with a score above 0.98, and they form the vast majority of identifications (see Figure~\ref{100cases-roc-vs-Nr}.B). 
Secondly, the smaller communities can be harder to detect, and this is more prominently seen for communities with size below 20 nodes. 

The previous results depend on the specific realization that generates the 20 tuples $(r, N_r)$, Fig~\ref{fig.100cases-roc}.(C), as well as the other random features. 
Furthermore, we'd like to explore how the mixing (relative magnitudes between $p$ and $q$) influences those results. 
So,  we  now employ a Monte Carlo strategy, for which the previous scenario can be considered one of its specific realizations. 
The steps are:

\begin{itemize}
\item[a.] We fix in inner and outer probabilities $(p,q)$. 
\item[b.] Matrix $\mathbf{X} = [\mathbf{A}_1~\mathbf{A}_2~...~\mathbf{A}_{H}]$ is generated by using steps (1-4).
\item[c.] The fraction of communities detected and planted is calculated as before. 
\item[d.] Steps (a-c) are repeated $N_{\rm MC} = 20$ times. 
\end{itemize}

\begin{figure}
	\centering
	\includegraphics[width=0.8\linewidth]{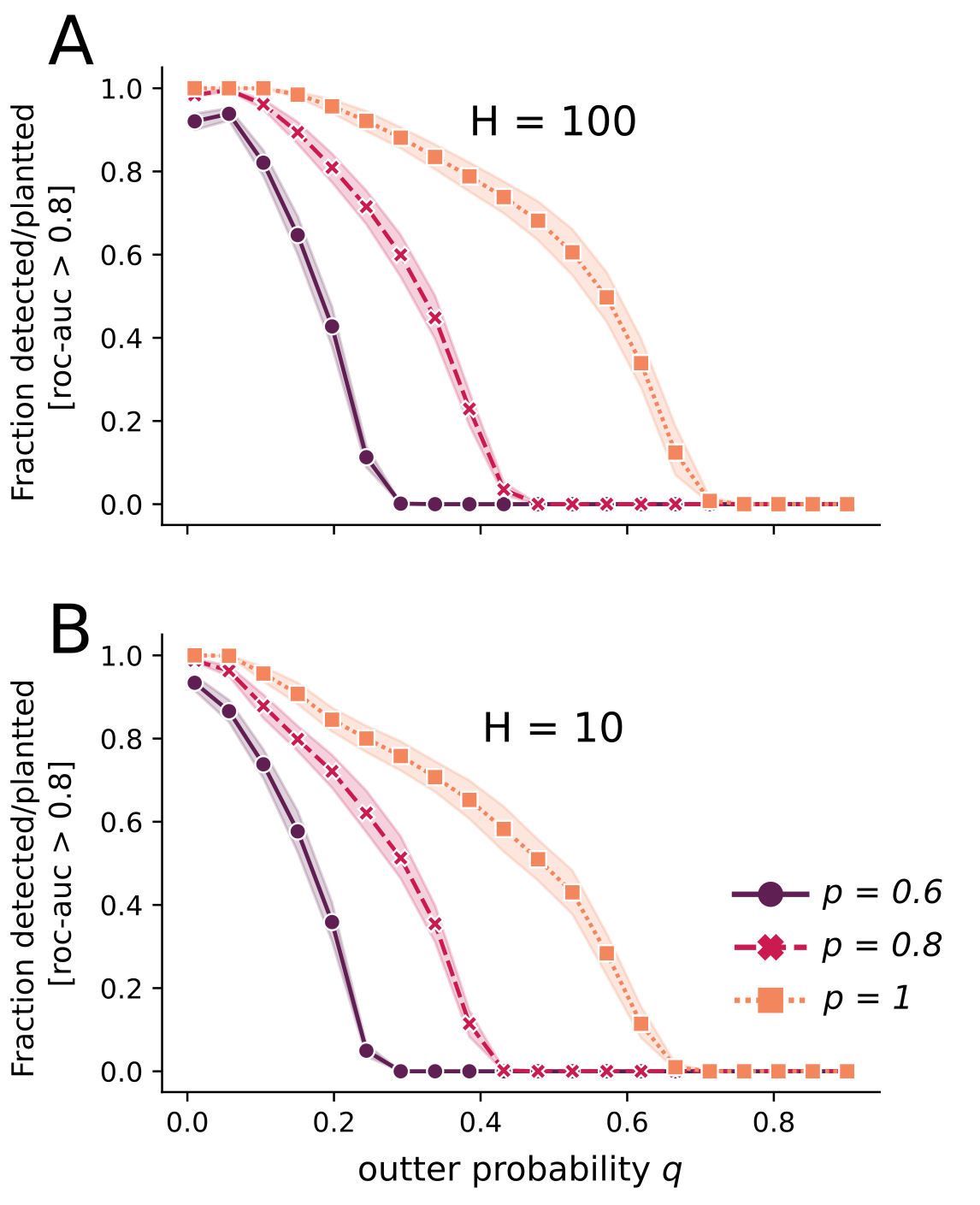}
	\caption{\label{fig.fraction-monte-carlo} Fraction of detected communities with ROC-AUC score above $0.8$.  
	 The curves show the mean ($\pm$ standard deviation) over 20 Monte Carlo runs for an increasing outer probability $q\in[0.01,0.9]$ and three fixed values of the inner probability $p$. 
	 The number of conditions is (A) $H=100$ (see Figs.~\ref{fig.100cases-factors}-\ref{fig.100cases-roc}) and (B) $H=10$, both for a network of size $N=2000$. 
	 A larger $H$ increases the chance of a given community appears on multiple conditions, which increases its chance of been detected. This causes the slower decay of the curves in (A) as compared to (B).  
	}
\end{figure}

That was done for 20 increasing values of the outer probability $q\in[0.01, 0.9]$. 
We considered scenarios with inner probability $p=0.6$, $0.8$ and 1 (1-cliques). 
Figure~\ref{fig.fraction-monte-carlo}.A shows the result (mean $\pm$ standard deviation) for $H=100$ conditions. 
It is seen that lower the mixing (i.e., $p$ been more prominent than $q$) the detection curve (i) decays slower and (ii) starts decaying at a larger $q$ value. 
As mentioned before, if a given small community appears in several conditions, its ``signal" would be stronger. That would counterbalance its small size, allowing it to be detected. 
Because of that, we repeat the experiment for a much lower number of conditions $H=10$. 
In this scenario, it is much less likely that any of the 20 possible communities will appear several times. 
As a consequence, Fig.~\ref{fig.fraction-monte-carlo}B, the detection curves start decaying at a lower value of $q$ as compared to the scenario with $H=100$. 
Regarding the speed of decay (slop), it is very similar to the previous case but for a small segment (between $q\approx.1$ and 0.2) for the $p=1$ curve.

\subsection{Genetic data}
\label{sec.real-2}
Here we give an illustrate example of how the CDR could be used to explore the differential structure of real-world data, in the context of differential gene co-expression network analysis. 
This is done assuming an exploratory data analysis point-of-view, and then showing evidence of the extent to which the CDR approach can recover subtle and nuanced information that is otherwise spread along with the original datasets.

We will use a tutorial dataset for single-cell analysis~\citep{Kang2017}. 
It consists of gene expression data of peripheral blood mononuclear cells (PBMCs) divided into two groups based on treatment:  one with and the other without interferon-$\beta$ (IFNB) stimulation. 
Both the dataset and tutorial (with R code) for its analysis can be found at \url{https://satijalab.org/seurat/v3.1/immune_alignment.html}, and within the  Seurat R package~\cite{seurat_pkg}.
The tutorial (and R package) contains the log-normalized expression data of 2000 genes for 12 different cell types. 
To illustrate the CDR, we arbitrarily selected two cell types: CD4 Naive T cell (henceforth refereed as T-cells) and CD14-mono cells, both labelled as STIM (IFNB stimulation) or CRTL (control) regarding their respective treatment group. 

Before proceeding with the CDR, we will build $H=4$ (co-expression) networks, which conceptually represent what we called conditions in the CDR method. 
Let the matrices $\mathbf{E}_h$, with $h=1, ..., H$, be the expression data with genes in rows and samples in columns. 
We will refer to these conditions as ${\rm Tc_{ctrl}}$, ${\rm Tc_{stim}}$, ${\rm cMono_{ctrl}}$ and ${\rm cMono_{stim}}$. 

Before continuing, three remarks are worth noticing. 
Firstly, 
we emphasize here that these two cell types were selected arbitrarily and exclusively to illustrate the use of the CDR in an exploratory analysis to highlight similar and dissimilar networks structures across different conditions: 
in this section, the focus is not on searching for any biological meaningfully result from that analysis. 
Secondly, 
in a real-world exploratory analysis, the selection of cell-types and experimental groups should be guided by the specific research questions that the {\it specialist}-subject researcher aims for. 
For example, the selection above could be justified if the interest is to search for the $2\times 2$ (cell-types $\times$ treatments) structural (dis)similarities between the co-expression networks of Tc and CD14-mono cells (i.e., between themselves and between treatment).  
So, {\it specific} research questions are the first things to consider before selecting the datasets to building the concatenated matrix $\mathbf{X}$. 
Finally, 
the estimate of a gene co-expression adjacency from expression data is a very active research topic {\it per se}~\cite{Choobdar2019,Liesecke2019}. 
It is not our aim to discuss, or provide any guidance, on what method would provide more reliable results. 
The application of the CDR and the illustrative goal of this section are independent of possible direct biological inference. However, it is expected that meaningful biological results will depend on how the estimated adjacency matrices are an accurate representation of the underlying reality. 
Therefore, this aspect is pertinent in a real-world application.

From the selected data, $H=4$ weighted adjacency networks were built as follows. 
One of the most often used methods in gene network analysis is the weighted gene correlation network analysis (WGCNA)~\cite{Langfelder2008}, which can be used for finding clusters (modules) of highly correlated genes. 
Its initial step (after standard prepossessing, already done in the case of the Seurat dataset) is to build a weighted network from the expression data from a {\it co-expression similarity} matrix $\mathbf{S}={\rm cor}(\mathbf{E})$ (the correlation between the expression of genes across samples). 
Then, the {\it weighted network adjacency} is defined as $\mathbf{W}=\mathbf{S}^{\beta}$. 
The parameter $\beta$ is chosen in a way to provide  $\mathbf{W}$  with a scale-free topology. 
The motivation is that real-world biological networks often show the scale-free topology, so an appropriate selection of $\beta$ would provide a more accurate representation of reality. 
Given that definition through correlations, a standard step is to remove genes with near to zero variance across samples. 
From the 2000 genes from the Seurat data, we removed the ones with a variance less than $10^{-10}$, yielding 1473 genes. 
From these, we select 1458 genes with the highest variance (corresponding to the 0.1 quantile). 
After that, the four expression data matrices sizes are  
$\mathbf{E}_1\in \mathbb{R}^{1458,1034}$, 
$\mathbf{E}_2\in \mathbb{R}^{1458,1579}$, 
$\mathbf{E}_3\in \mathbb{R}^{1458,1036}$ and 
$\mathbf{E}_4\in \mathbb{R}^{1458,3285}$. 

Because we want to focus on the information that the application of the CDR method {\it alone} can provide, we will {\it not} make use of such enhanced representation of the adjacency matrix as provided by WGCNA. 
We will set $\beta=1$, or equivalently use the similarity matrix $\mathbf{S}$ itself. 
Specifically, the four weighted network adjacency matrices will be
$\mathbf{W}_h = {\rm cor}(\mathbf{E}_h)$ with $h=1, ..., 4$ 
(for the conditions ${\rm Tc_{ctrl}}$, ${\rm Tc_{stim}}$, ${\rm cMono_{ctrl}}$ and ${\rm cMono_{stim}}$, respectively). 
Note that each of these four matrices has size $1458 \times 1458$, with nodes corresponding to genes. 
To decrease the computational time of the eigendecomposition, all correlations between $\pm 0.02$ were set to zero. 
That procedure increases the sparsity of $\mathbf{X}$, allowing more efficient use of algorithms for the decomposition of sparse matrices. 
Finally, the input for the CDR is the matrix $\mathbf{X} = [\mathbf{W}_1~\mathbf{W}_2~\mathbf{W}_3~\mathbf{W}_4]$ of size $1458 \times 5832$.

Now we illustrate an exploratory analysis with the CDR.  
The main goal is to demonstrate that the information provided by the rotated factor loadings reflects the structure encoded in the adjacency matrices $\mathbf{W}_h$. 
The number of varimax rotated vectors was 60. 
We experimented with different values (e.g., 40 and 100), with no impact on the following results.

\begin{figure*}[!]
	\centering
	\includegraphics[width=1\linewidth]{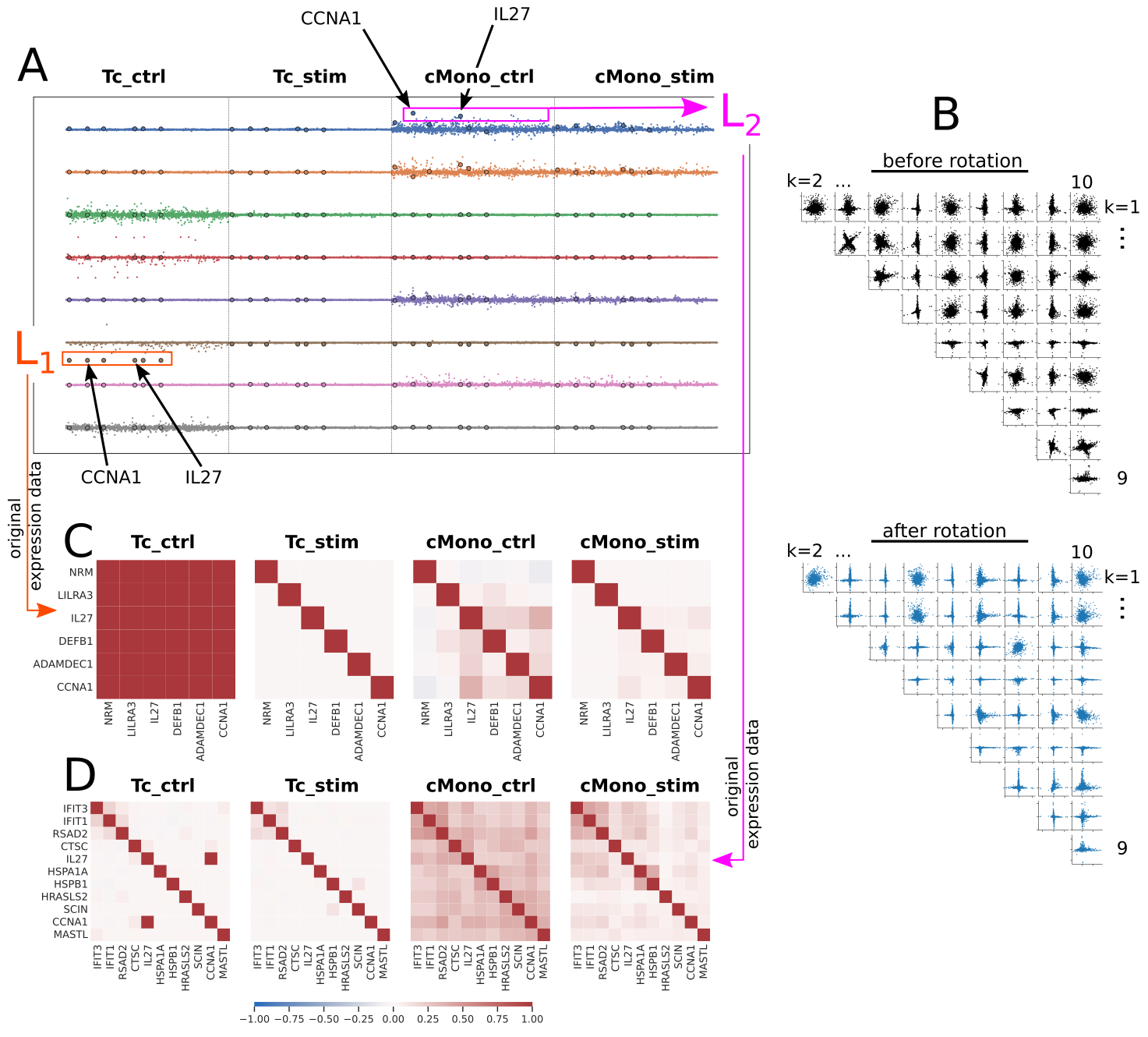}
	\caption{\label{fig-gene-unified}
	Applying the CDR framework to gene co-expression networks. 
	(A) Cell and/or group specific nuances are clearly suggested by the rotated loading vectors $\hat{\mathbf{\Lambda}}^*_{\bullet k}$. 
	(B) Scatter plots $\hat{\mathbf{\Lambda}}^{*}_{\bullet k_{1}}$ {\it versus} $\hat{\mathbf{\Lambda}}^*_{\bullet k_{2}}$ (lower panel), with $k=1,\ldots, 10$ and $ k_{1} \neq  k_{2}$, show the enhanced interpretability (i.e., simple structure) provided by the varimax rotation as compared to the unrotated $\hat{\mathbf{\Lambda}}_{\bullet k}$ (top).
	(C) Correlations in the original expression data for the selected genes in the set ${\rm L_1}$ show that they form a highly connected community {\em specific} for Tc cells in the control group. 
	(D) As before, but for the genes from the set ${\rm L_2}$. 
	These genes are part of a more complex structure of highly connected nodes specific for cMono cells, and more ``stronger" in the control group than in the stimulated one. }
\end{figure*}

The leading 8 rotated factor loading vectors $\hat{\mathbf{\Lambda}}^*_{\bullet k}$ are shown in Fig.~\ref{fig-gene-unified}A. 
We will sometimes refer to them simply by their indexes, $k$. 
A simple visual inspection of the magnitudes across conditions (i.e., the four segments of length 1458) allows one to {\em interpret} that for 

 \begin{enumerate}
     \item $k=1$, 2 and 5, segments 3 and 4 reflect a structure peculiar to cMono cells, which is stronger (in ``a sense", more on this later) in the control group than in the stimulated one. 
      \item  $k=7$, those segments 3 and 4 reflect again a structure peculiar to cMono cells, but which is now equivalent in both stimulated and control groups. 
     \item  $k=3$, 4, 6 and 8, the first segment reflects a structure related only to Tc cells in the control group. 
 \end{enumerate}
Therefore, we will use that list of statements above as a metaphorical ``navigation map" to guide the creation of some hypotheses, which can be tested by using the original datasets $\mathbf{E}_h$. 
But before that, we will explore some complementary information provided by the scatter plots between the leading $k=1, ..., 10$ vectors, Fig.~\ref{fig-gene-unified}B. 
Note that because consecutive nodes are not within the same (differential) community structure, the effects of the rotation cannot be seen by plotting the factor loading vectors before and after rotation, as done the case of the synthetic network shown in Fig.~\ref{fig.100cases-factors}. 
This makes the scatter plots a suitable alternative. 
Firstly, consider the effects of the varimax rotation in disentangling the information between the vectors. 
For instance, the rotation of a cross-shaped structure can be seen for $(k_1, k_j) = (2,3)$ (second row, third column of Fig.~\ref{fig-gene-unified}B top and bottom panels), meaning that the factor loading vectors $\hat{\mathbf{\Lambda}}^*_{\bullet 2}$ and $\hat{\mathbf{\Lambda}}^*_{\bullet 3}$ become uncorrelated after rotation --- 
and because of that, they should reflect cell-group specific and distinct structures encoded into $\mathbf{X}$: a feature that agrees with the statements (1) and (3) above. 
Another effect is seen as the reduced number of circular cloud of points, which after rotation are replaced by linear structures parallel to the horizontal and or vertical axis. 
Secondly, some scatter plots remain as circular clouds even after rotation. 
For instance, the features captured by vectors $k=1$ and 2, which are specific to cMono cells, are correlated --- this suggests that those vectors could tell different nuances of the same differential structure that make those cells apart from the Tc cells. 
So, the inspection of the scatter plots provides complementary information for the factor loading vectors interpretability.

Taken together, the information above provides guidance for the creation and exploration of a vast amount of hypotheses. 
Next, we will develop and explore some possibilities, which we believe provide the most complete illustrative example of the power of an exploratory analysis through the CDR. 
Note that, in practice, this should be done side-by-side with an expert in the field related to the dataset.

Consider the $k=6$ first segment, which should refer to a specific structure for Tc cells in the control group. 
Let ${\rm L_1}$ be the set of the six genes associated with the nodes with the largest (absolute) factor loadings. 
That selection is highlighted with a rectangular (orange) box in Fig.~\ref{fig-gene-unified}A, and the genes are ${\rm L_1}=$\{ADAMDEC1, CCNA1, DEFB1, IL27, LILRA3, NRM\}. 
Larger marker size was used for all occurrences of those six genes, which allows one to verify that they appear (collectively) with a large loading only at the $k=6$ first segment. 
However, the genes CCNA1 and IL27 appear with a large loading in $k=1$ segment 3, a condition that should reflect a structure {\em specific} for cMono cells in the control group. 
Based on what we have discussed in the example of the synthetic data (Sec.~\ref{sec.synthetic}), those observations suggest the following inter-connected hypotheses:
\begin{enumerate}[I.]
    \item  The ${\rm L_1}$ genes form a highly connected community, which is specific for the Tc cells in the control group. 
    \item  That community is present {\it solely} for the Tc cells in the control group.
    \item  There is another structure ${\rm L_2}$ (see the purple rectangular selection in Fig~\ref{fig-gene-unified}A) specific for cMono cells in the control group, for which genes CCNA1 and IL27 are members. 
    \item  The structural role of genes CCNA1 and IL27 within community ${\rm L_1}$ is different from their structural roles in ${\rm L_2}$. 
\end{enumerate}

\noindent
\begin{remark}
The set  ${\rm L_2}$ contains eleven genes: IFIT3, IFIT1, RSAD2, CTSC, IL27, HSPA1A, HSPB1, HRASLS2, SCIN, CCNA1, MASTL. 
\end{remark}

Those hypotheses can be tested by estimating the strength of co-expression (correlations) of the genes sets ${\rm L_1}$ and ${\rm L_2}$ within each {\em original} expression dataset $\mathbf{E}_h$. 
Figures~\ref{fig-gene-unified}C-D show the correlations for ${\rm L_1}$ and ${\rm L_2}$, respectively, which strongly agree with the hypotheses above. 
Specifically, the following features can be seen. 
(i) The six ${\rm L_1}$ genes have a highly-correlated co-expression {\it solely} for the Tc cells in the control group, hence forming a differential community. 
(ii) There is a slightly higher correlation between genes CCNA1 and IL27 for the cMono in the control group, Figure~\ref{fig-gene-unified}C, showing that these two genes can have a different structural role in another community (related specifically to cMono in the control group), and which we will track now. 
To allow better visualization of the community structure, the four heat-maps in  Fig.~\ref{fig-gene-unified}.D were plotted after rearranging their columns and rows based on the hierarchical clustering dendrogram (HCD) of the cMono control group.  
In the third heat-map of Fig.~\ref{fig-gene-unified}.D, (iii) the eleven genes in ${\rm L_2}$ show a somewhat more complex community structure (i.e., 3 or 4 small communities, without a crisp frontier, forming a larger community). 
As suggested by the ordering provided by the HCD, the co-expression between genes  CCNA1 and IL27 forms a link between two different communities (one of them formed by the last 4 genes). 

Nevertheless, the last two heat-maps in Fig.~\ref{fig-gene-unified}D can be used to test another hypothesis. 
Previously, we mentioned that for $k=1$ the segments 3 and 4 should reflect not only a structure more peculiar to the cMono cells but that that structure should be stronger (in a sense) in the control group than in the stimulated one. 
Figure~\ref{fig-gene-unified}.D shows that this is indeed the case and that the structure is stronger in {\em two} senses: 
(a) there are links between the sub-communities in the control group, but they are broken in the stimulated group (for which the heat-map suggests the existence of two sub-communities, one with genes IFIT3, IFIT1, RSAD2m, and the other with genes HSPA1A, HSPB1); 
(b) the third sub-community in the control group (with genes HRASLS2, SCIN, CCNA1, MASTL) disappears in the stimulated group. 

Taken together, those results support the conclusion that the ``story-telling"-like information provided by the CDR approach can guide one to recover subtle and nuanced information that is otherwise spread along with the original datasets.


\section{Conclusion}
\label{sec.conc}
In this paper, we have proposed a theory and method (CDR) for the characterization of shared ``structural roles" of nodes {\em simultaneously} within and between networks, whose outcome is a highly interpretable map. 
Without loss of generality, for the sake of presentation we have assumed that each network represents different experimental conditions. 
They could be, equivalently, representations of a time-evolving network (i.e., in different time-points), layers in a multiplex network etc. 

Supported by a simple and transparent theory, rooted in the factor analysis framework, the method provides flexibility to address different research-field specific questions. 
This is accomplished by defining {\em what} is the scientific-meaningful characteristic (or relevant feature) of a node at the problem at hand, and then mapping it to an appropriate mathematical similarity construct to estimate the proximity from measured data. 
In the context of differential network analysis, with communities of highly connected nodes, in this paper the method was illustrated by assuming as the relevant feature the similarity of the list of neighbours between nodes, captured by the notion of proximity through the inner (vector) product.

The insights provided by the method and its accuracy have been explored in numerical benchmarks generated by a stochastic block model. 
In the scope of non-overlapping communities (which could be present or absent on a given condition) the results have shown the method's high accuracy despite very different (i) community sizes, (ii) total number of communities within a given condition and (iii) number of networks been compared (i.e., experimental conditions). 

In particular, we have argued that the strength of the method is its ``story-telling"-like characterization of the differential network structure encoded in a set of networks. 
This was illustrated with an exploratory analysis of gene expression datasets, a context called differential co-expression network analysis. 
In particular, applying the CDR to the co-expression networks from two cell types $\times$ two treatments allowed us to elaborate hypothesis related to {\em unique} and {\em subtle} nuances of the structural differences between the networks. 
Those hypotheses were then tested and confirmed by the original dataset. 

Our vision is of a virtuous cycle, where the CDR is used as a hypothesis generator by pinpointing unexpected differential structures within the data, leading to further investigations and providing new insights. 

The analysis of genetic data was used, as well, to illustrate how the fundamental underlying ideas of the CDR could be applied over and above the other methods that have been developed within different scientific communities. 
For this, it has been applied on (a simplified version of) weighted gene correlation network analysis. Notwithstanding, due to (i) the CDR flexibility and (ii) the several ways the theoretical basis of factor analysis allows one to extract the factor loadings (here we have chosen the PCA), those ideas could even be used as a bridge between the several methods of network analysis (from different communities) to nurture a deeper theoretical understanding. 
These and other possibilities we leave for future work.

\section*{ACKNOWLEDGMENTS} 
\noindent
LLP and MS acknowledge financial support from Australian Research Council Discovery Grant (DP 180100718). 
This work was partly supported by the Pawsey Supercomputing Centre with funding from the Australian Government and the Government of Western Australia.


\pagebreak

%

\end{document}